\documentclass[a4paper,useAMS,usenatbib,usegraphicx]{mn2eadapt}

\usepackage[latin1]{inputenc}
\usepackage{amsmath}
\usepackage{amsfonts}
\usepackage{amssymb}
\usepackage[english]{babel}

\usepackage{times}
\usepackage{color}
\usepackage[colorlinks=false, pdfborder={0 0 0},dvips]{hyperref}

\newcommand{\Lsun}{L_{\odot}}
\newcommand{\kms}{km~s$^{-1}$}

\newcommand{\NaI}{Na~{\sc i}}

\newcommand{\SII}{S~{\sc ii}}
\newcommand{\SIII}{S~{\sc iii}}

\newcommand{\SiII}{Si~{\sc ii}}
\newcommand{\SiIII}{Si~{\sc iii}}

\newcommand{\TiII}{Ti~{\sc ii}}

\newcommand{\CoII}{Co~{\sc ii}}

\newcommand{\NiII}{Ni~{\sc ii}}
\newcommand{\Fefs}{$^{56}$Fe}
\newcommand{\Cofs}{$^{56}$Co}
\newcommand{\Nifs}{$^{56}$Ni}

\newcommand{\Dm}{$\Delta m_{15}(B)$}

\newcommand{\aap}{A\&A}
\newcommand{\mnras}{MNRAS}
\newcommand{\apj}{ApJ}

\newcommand{\apjl}{ApJ}
\newcommand{\aj}{AJ}

\newcommand{\sci}{Sci}

\begin{document}

\title[SN Ia line strength - luminosity correlations]{Spectral luminosity indicators in SNe Ia - \\Understanding the $\mathcal{R}$(\SiII) line strength ratio and beyond.}

\author[Hachinger et al.]{Stephan Hachinger$^{1}$,
Paolo A. Mazzali$^{1,2}$, Masaomi Tanaka$^{3}$, Wolfgang Hillebrandt$^{1}$,\protect\vspace{0.2cm}\\{\upshape\LARGE Stefano Benetti$^{2}$}\\
$^1$Max-Planck-Institut f\"ur Astrophysik, Karl-Schwarzschild-Str.\ 1, 85748 Garching, Germany\\
$^2$Istituto Nazionale di Astrofisica-OAPd, vicolo dell'Osservatorio 5, 35122 Padova, Italy\\
$^3$Department of Astronomy, Graduate School of Science, University of Tokyo, Hongo 7-3-1, Bunkyo-ku, Tokyo 113-0003, Japan}

\date{arXiv v1, 2008-06-25.}

\pubyear{2008}
\volume{}
\pagerange{}

\maketitle

\defcitealias{hac06}{Paper~I}

\begin{abstract}

SNe Ia are good distance indicators because the shape of their light curves,
which can be measured independently of distance, varies smoothly with
luminosity. This suggests that SNe Ia are a single family of events. Similar
correlations are observed between luminosity and spectral properties. In
particular, the ratio of the strengths of the \SiII\ $\lambda5972$ and
$\lambda6355$ lines, known as $\mathcal{R}$(\SiII), was suggested as a potential
luminosity indicator. Here, the physical reasons for the observed correlation are
investigated. A Monte-Carlo code is used to construct a sequence of synthetic
spectra resembling those of SNe with different luminosities near $B$
maximum. The influence of abundances and of ionisation and excitation 
conditions on the synthetic spectral features is investigated. The ratio 
$\mathcal{R}$(\SiII) depends essentially
on the strength of \SiII\ $\lambda5972$, because \SiII\ $\lambda6355$ is
saturated. In less luminous objects, \SiII\ $\lambda5972$ is stronger because of
a rapidly increasing \SiII/\SiIII\ ratio. Thus, the correlation between 
$\mathcal{R}$(\SiII) and luminosity is the effect of ionisation balance. The
\SiII\ $\lambda5972$ line itself may be the best spectroscopic luminosity
indicator for SNe Ia, but all indicators discussed show scatter which
may be related to abundance distributions.

\end{abstract}


\section{Introduction}

Type Ia supernovae (SNe Ia) form a relatively homogeneous supernova subclass.
They supposedly originate from accreting C-O 
white dwarfs that undergo explosive burning as they approach the 
Chandrasekhar mass limit \citep{hil00}. Their enormous light output 
is powered by radioactivity of synthesised \Nifs\ and of its daughter nuclide \Cofs\
\citep{arn82}.

Differences in the synthesised \Nifs\ 
mass result in a spread in luminosity of $\sim\!\!3$ mag in the $B$ band 
\citep{maz01b,maz07}. \citet{phi93} found that the decline in $B$-band 
magnitude within 15 days after $B$ maximum, $\Delta m_{15}(B)$, 
correlates inversely with the $B$-band maximum luminosity
$M_{B,\textrm{max}}$ and thus can be used as a luminosity indicator. Thus, SNe
Ia can be used as distance indicators in cosmological studies (``standard candles'').

Optical spectra reflect in detail the variations among SNe Ia. In the photospheric phase,
which lasts until a few weeks after maximum light, P-Cygni features (blueshifted
absorption plus less prominent emission centred at the rest wavelength) are
superimposed on a ``pseudo-continuum'' \citep{pau96}. The line positions and
strengths depend on the abundances and on the excitation and 
ionisation conditions within the debris.  \citet{nug95} showed that
the ratio of the depths of two \SiII\ lines ($\lambda 5972$ and $\lambda 6355$)
at $B$-band maximum, $\mathcal{R}$(\SiII), is larger for dimmer SNe, and
suggested that this is due to temperature differences. Accordingly,
$\mathcal{R}$(\SiII) could be used as a distance-independent spectroscopic
luminosity indicator \citep[e.g.][]{bon06}. Other spectral features have subsequently
been suggested as possible luminosity indicators (\citealt{hac06},
hereafter \citetalias{hac06}; \citealt{bon06}; \citealt{fol08}). Still, the
detailed physics behind the $\cal R$(\SiII)-$L$ correlation have not been
explained.

This is the purpose of this work, where we explain the $\cal R$(\SiII)-$L$
correlation investigating the details of the connection between the physics of
the envelope and line strengths. We present a sequence of spectral models
calculated with a Monte-Carlo (MC) radiative transfer code. The synthetic
spectra resemble real SNe with different luminosities shortly after $B$ max. In
the model atmospheres, we analyse the conditions which influence the line
strengths entering $\cal R$(\SiII).

In the following, first we describe the concept of the synthetic spectral
sequence (Sec. 2). In Sec. 3, we show that our models reproduce the $\cal
R$(\SiII)-$L$ correlation. In Sec. 4 we explain the correlation analysing the
details of the  formation of the lines contributing. After discussing spectral luminosity
indicators and other applications of line-strength measurements (Sec. 5), we give
conclusions (Sec. 6).

\section{Concept of the model sequence}
\label{sec:modelsequence}

\subsection{Model SN envelope and radiative transfer simulation}

Our work makes use of a 1D Monte Carlo (MC) radiative transfer code
(\citealt{abb85}, \citealt{maz93a}, \citealt{luc99} and \citealt{maz00}, which
describes the version used in this work) to compute synthetic SN spectra. The
code has successfully been applied to many SNe (SN~1990N: \citealt{maz93b},
etc.). 

The code computes the radiative transfer through the SN above an assumed 
photosphere. The envelope is assumed to expand homologously (e.g. 
\citealt{roe05}), meaning that the ejecta move radially with 
$r=v\cdot t$, where $r$ is the distance from
the centre, $t$ the time from explosion and $v$ the velocity. Radius and
velocity can therefore be used interchangeably for each gas particle. 
The density structure is taken from a standard explosion model (W7,
\citealt{nom84}), scaled according to the time $t$. The chemical composition of
the model atmosphere can be freely adjusted, but is assumed to be homogeneous in
radius. 

From the photosphere, which is located at an adjustable $v_{\textrm{ph}}$,
``packets'' of thermal radiation [$I_{\nu}^{+}=B_{\nu}(T_{\textrm{ph}})$] 
are emitted into the atmosphere. As they propagate through the 
atmosphere, radiation packets can undergo Thomson
scattering on electrons or line excitation-deexcitation processes, which are
treated in the Sobolev approximation. The process of photon branching is
included, which implies that the transitions for excitation and deexcitation
can be different. In all processes, radiative equilibrium is conserved. Taking
the backscattering into account, the code matches a given output luminosity
$L$.

The action of the radiation field onto the gas is calculated using a modified
nebular approximation, which mimics the effects of NLTE. The envelope is
discretised into 80 shells. For each shell, a radiation temperature $T_R$ and a
dilution factor $W$ are calculated. These quantities mostly determine the
excitation and ionisation state.

The code iterates the radiation field and
the gas conditions. After convergence, the output spectrum is obtained 
from a formal integral solution of the transfer 
equation \citep{luc99}.

\subsection{Sequence of synthetic spectra}
\label{sec:parameterchanges}

\begin{figure}   
   \centering
   \includegraphics[width=8.4cm]{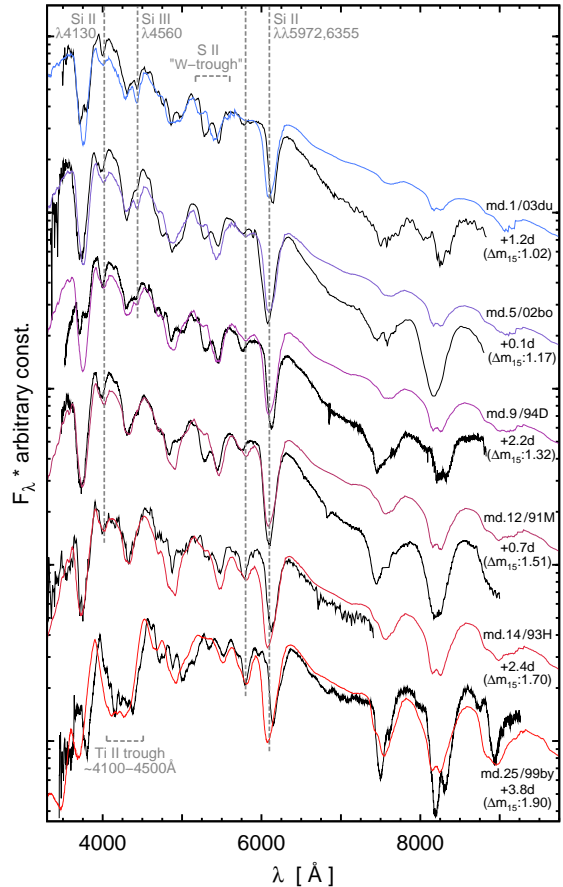}
   \caption{Selected spectra from the model sequence, and observed counterparts (black) with \Dm\ values (see \citetalias{hac06}). All spectra are plotted logarithmically and have been adapted to the total reddening of SN~2003du. Approximate identifications are given for features relevant for this work.}
   \label{fig:spectra}
\end{figure}

For our study we computed a sequence of 25 spectral models. The start and end
points of the sequence were chosen to match a luminous SN Ia (model no. 1:
SN~2003du, \Dm$=1.02$, $M_{B\textrm{,model}}=-19.16$) and a dim one (model no.
25: SN~1999by, \Dm$=1.90$, $M_{B\textrm{,model}}=-16.86$), respectively, close to
$B$ maximum. Observed and synthetic spectra are shown in Fig. \ref{fig:spectra};
further data and references are given in Table 1. 

\begin{figure}
   \centering
   \includegraphics[angle=270,width=8.4cm]{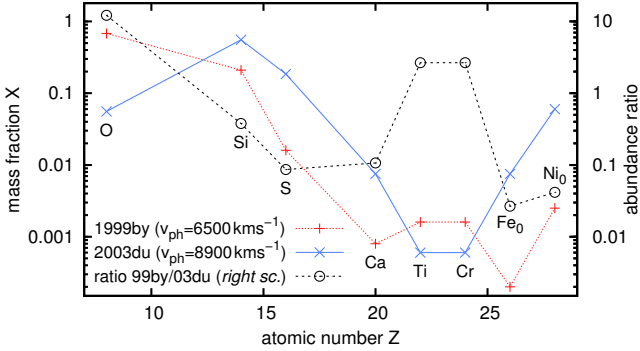}
   \caption{Abundances of the 99by and 03 du model. The Fe and Ni abundances given are $X$(Fe)${}_0$ and $X$(${}^{56}$Ni)${}_0$ (see tab. \ref{tab:modelparameters}). The SN~99by model is generally characterised by very small abundances of burning products, except for relatively large amounts of Ti and Cr (Fig. \ref{fig:abundances}). Despite the decrease in Si abundance, \SiII\ $\lambda5972$ is stronger in dimmer objects.}
   \label{fig:abundances}
\end{figure}

The free model parameters shown in Table \ref{tab:modelparameters} and Fig.
\ref{fig:abundances} were iterated to optimise the fit. Some deviations between
models and data remain, but they have no influence on our investigation. In the
luminous models there is an excess flux in the red caused by the assumption of a
blackbody lower boundary. Dim models show a mismatch in line velocities, which
should vanish with a refined density and abundance stratification (cf.
\citealt{tau08}).

\begin{table}
\scriptsize
\caption{First (no. 1) and last (no. 25) model of the sequence: Data of the objects and spectra used for modelling.}
\label{tab:obsparameters}
\centering
\begin{tabular}{cccccccc}
 $\!\!\!$Object/epoch/ $\!\!\!\!$& $\!\!\!\!\Delta m_{15}\!\!\!\!$ & Total reddening & $\!\!\!$Distance modulus$\!\!\!$ & $\!\!\!\!\!\!\!$Redshift$\!\!\!\!\!\!\!$ & References \\
 model no. & [mag] & $\!\!\!\!\!\!E(B-V)^{\textrm{a)}}$  [mag]$\!\!\!\!\!\!$ & $\mu$ [mag]     & $\!\!$ $z$ $\!\!$ & for SN,$\mu$,$z$\\ \hline
$\!\!\!$03du/+1.2d/\#1$\ \ \!\!\!\!\!$& 1.02 & 0.01 & 32.79 &$\!\!$ 0.0068 $\!\!$& $\!\!\!$ \citet{sta07}$\!\!$ \\
$\!\!\!$99by/+3.9d/\#25$\!\!\!\!\!$& 1.90 & 0.01 & 30.75 &$\!\!$ 0.0021 $\!\!$& $\!\!\!\!$\citet{gar04}${}^{\textrm{b)}}\!\!\!\!$\\
\end{tabular} \newline
\begin{flushleft}
${}^{\textrm{a)}}$ The values given represent the galactic extinction $E(B-V)_{\textrm{gal}}$, as the host galaxy extinction is approximately zero for both SNe (see references). $E(B-V)_{\textrm{gal}}$: from NED (see acknowledgements); we averaged the \citet{bur82} and \citet{sch98} values.\newline
${}^{\textrm{b)}}$ Spectrum combined from original spectra with different grating setups (taken between JD 2451312.64 and 2451312.66, available at http://www.cfa.harvard.edu/supernova/) and fine-calibrated to match photometry by multiplying with a linear function (using \textsc{IRAF} and \textsc{SYNPHOT} routines/data, see acknowledgements).
\end{flushleft}
\end{table} 

\begin{table*}
\scriptsize
\caption{Overview of the final parameters for models no. 1 and 25. Mass fractions less significant have been omitted.}
\label{tab:modelparameters}
\centering
\begin{tabular}{lllllllllllllc}
Model	$\!\!\!\!$ & $\!\!$	$\lg\left(\frac{L}{\Lsun}\right)$\vspace{-0.2cm}	$\!\!\!\!$ & $\!\!\!\!$	$v_{\textrm{ph}}$	$\!\!\!\!$ & $\!\!\!\!$	$t^a$	$\!\!$ & 	\multicolumn{9}{c}{Element abundances (mass fractions) }  \\						\vspace{-0.45pt} & \vspace{-0.45pt} & \vspace{-0.45pt} & \vspace{-0.45pt} & \vspace{-0.45pt} & \vspace{-0.45pt} & \vspace{-0.45pt} & \vspace{-0.45pt} & \vspace{-0.45pt} & \vspace{-0.45pt} & \vspace{-0.45pt} & \vspace{-0.45pt} & \vspace{-0.45pt}  \\
$\!\!\!\!$ &  $\!\!$		$\!\!\!\!$ & $\!\!\!\!$	[\kms]	$\!\!\!\!$ & $\!\!\!\!$	[d]	$\!\!$  & $\!\!\!\!$	$X$(O)	$\!\!\!\!$ & $\!\!\!\!$	$X$(Mg)	$\!\!\!\!$ & $\!\!\!\!$	$X$(Si)	$\!\!\!\!$ & $\!\!\!\!$	$X$(S)	$\!\!\!\!$ & $\!\!\!\!$	$X$(Ca)	$\!\!\!\!$ & $\!\!\!\!$	$X$(Ti)	$\!\!\!\!$ & $\!\!\!\!$	$X$(Cr)	$\!\!\!\!$ & $\!\!\!\!$	$X$(Fe)$_{0}{}^{\textrm{b)}}$	$\!\!\!\!$ & $\!\!\!\!$	$X$(\Nifs)$_{0}{}^{\textrm{b)}}$ \\ \hline
no. \ 1: 03du/+1.2 $\!\!\!\!$ & $\!\!$	9.61	$\!\!\!\!$ & $\!\!\!\!$	8900	$\!\!\!\!$ & $\!\!\!\!$	21.0	$\!\!$  & $\!\!\!\!$	0.056	$\!\!\!\!$ & $\!\!\!\!$	0.13	$\!\!\!\!$ & $\!\!\!\!$	0.56	$\!\!\!\!$ & $\!\!\!\!$	0.19	$\!\!\!\!$ & $\!\!\!\!$	0.0075	$\!\!\!\!$ & $\!\!\!\!$	0.0006 $\!\!\!\!$ & $\!\!\!\!$0.0006 $\!\!\!\!$ & $\!\!\!\!$	0.0075	$\!\!\!\!$ & $\!\!\!\!$	0.060	\\
no. 25: 99by/+3.8 $\!\!\!\!$ & $\!\!$	8.82	$\!\!\!\!$ & $\!\!\!\!$	6500	$\!\!\!\!$ & $\!\!\!\!$	21.0	$\!\!$  & $\!\!\!\!$	0.68	$\!\!\!\!$ & $\!\!\!\!$	0.085	$\!\!\!\!$ & $\!\!\!\!$	0.21	$\!\!\!\!$ & $\!\!\!\!$	0.016	$\!\!\!\!$ & $\!\!\!\!$	0.0008	$\!\!\!\!$ & $\!\!\!\!$	0.0016	$\!\!\!\!$ & $\!\!\!\!$	0.0016	$\!\!\!\!$ & $\!\!\!\!$ 0.0002 $\!\!\!\!$ & $\!\!\!\!$	0.0025	
\end{tabular} 
\newline
\begin{flushleft}
${}^{\textrm{a)}}$ Time from explosion onset. The rise times for 03du and 99by are assumed to be 19.8d and 17.1d, respectively. This complies with observational studies \citep{rie99,con06} and leads to a similar time from explosion of $\sim\!\!21.0$d for both models.\newline
${}^{\textrm{b)}}$ The abundances of Fe, Co and Ni in our models are assumed to be the sum of \Nifs\ and its decay chain products (\Cofs\ and \Fefs) on the one hand, and directly synthesised / progenitor Fe on the other hand. Thus, they are conveniently given in terms of the \Nifs\ mass fraction at $t=0$ [$X($\Nifs$)_0$], the Fe abundance at $t=0$ [$X(\textrm{Fe})_0$], and the time from explosion onset $t$.
\end{flushleft}
\end{table*} 

\begin{figure}   
   \centering
   \includegraphics[angle=270,width=8.4cm]{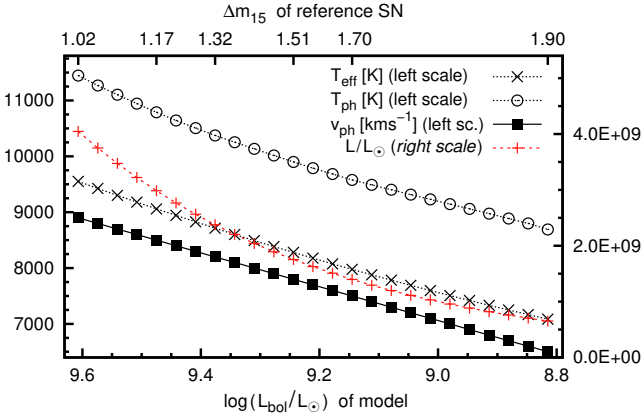}
   \caption{Input parameters $L_\textrm{bol}$ and $v_\textrm{ph}$ of our models, and resulting effective temperature $T_\textrm{eff}$. The photospheric blackbody temperature $T_\textrm{ph}$ after convergence is given for comparison.}
   \label{fig:efftemperatures}
\end{figure}

The focus of our fits is on obtaining a reasonable temperature structure. 
The decrease of temperature from luminous to dim SNe not only implies that
the spectra appear redder, but also influences line strengths and $\cal R$(\SiII) via excitation and 
ionisation. Choosing appropriate 
models for SNe~03du and 99by as the ends of 
the sequence, we ensure that we explore a realistic range of temperatures. 
The temperature structure of our models can to some degree be influenced by the 
choice of photospheric velocity. We fitted the pseudo-continuum of both SNe as 
well as possible. We did not increase temperatures further for SN~03du, because 
this would make \SiIII\ $\lambda4560$ too strong. The resulting effective 
temperatures (cf. \citealt{nug95})
\[
T_{\textrm{eff}} = \frac{ L^{1/4} }{(4\pi\sigma_B)^{1/4} (v_{\textrm{ph}}\cdot t)^{1/2}}
\]
($\sigma_B$: Stefan-Boltzmann's constant; $t\approx 21.0\textrm{d}$: time from 
explosion) of our models are shown in Fig. \ref{fig:efftemperatures}. For comparison
we also give the photospheric temperatures $T_\textrm{ph}$ which are calculated
by the code, taking the backscattering into account.

The other models of the sequence (no. 2-24) were calculated decreasing $\log L$
in constant steps. This leads to a relatively uniform distribution of models in
$M_B$. The photospheric velocity and all element abundances were changed by
constant amounts from one model to the next. An exception is the sulphur
abundance: in order to match the evolution of the \SII\ ``W-trough'' better, it
decreases with a larger slope among luminous SNe and with a smaller slope among
dim ones.

We assign $\Delta m_{15}(B)$ values to the models no. 1 and 25 and to four
intermediate ones. In order crudely to assign \Dm\ values to intermediate models, we
created ``optimal-fit'' models of the SNe 2002bo [\Dm$=1.17$], 1994D
[\Dm$=1.32$], 1991M [\Dm$=1.51$] and 1993H [\Dm$=1.70$] as reference models. We
then compared the sequence models to these by calculating a weighted sum of
quadratic deviations in the input parameters ($L_\textrm{bol}$, $v_\textrm{ph}$
and abundances), where the deviation in $L_\textrm{bol}$ was given the strongest
weight. The \Dm\ values of the reference models were then assigned to the four
sequence models that gave the closest match.

Fig. \ref{fig:spectra} shows the spectral models assigned a $\Delta m_{15}(B)$
value with their observed counterparts (SN~2002bo: \citealt{ben04}, SN~1994D:
\citealt{pat96}, SN~1991M: Asiago archive, SN~1993H: CTIO archive, with kind
permission of M.~M. Phillips).

\section{R(Si\textsc{ II}) in the model sequence}

\subsection{Line-strength measurements}

In order to compare models and observations, we measured the strengths of 
\SiII\ $\lambda 5972$ and \SiII\ $\lambda 6355$, and $\mathcal{R}$(\SiII) in
synthetic and observed spectra.

To quantify the line strength, the flux level inside a feature is compared to an estimated 
``pseudo-continuum'', which we define as a linear function approximately 
tangential to the spectrum at the edges of the feature. We then measure the
line strength as fractional depth:

\begin{equation}
F\:\!\!D(\lambda_{F\:\!\!D}) := \frac{F_{PC}(\lambda_{F\:\!\!D})-F(\lambda_{F\:\!\!D})}{F_{PC}(\lambda_{F\:\!\!D})}\textrm{,}
\end{equation}

where $F(\lambda_{F\:\!\!D})$ and $F_{PC}(\lambda_{F\:\!\!D})$ are the actual
flux and the pseudo-continuum, respectively, at a wavelength
$\lambda_{F\:\!\!D}$, which is chosen so as to obtain the maximum
{\usefont{OML}{cmm}{m}{it} FD}\footnote{\citet{nug95} quantified line strength
as ``depth'', i.e. the absolute difference between the pseudo-continuum and the
actual flux. The original $\mathcal{R}$(\SiII) depth ratio of \SiII\
$\lambda\lambda5972,6355$ behaves like the corresponding
{\usefont{OML}{cmm}{m}{it} FD} ratio in observations.}. To avoid the influence 
of noise, measurements were made in spectra convolved with a normalised
Gaussian function (FWHM: 15\AA).

\subsection{Evolution of {\usefont{OML}{cmm}{m}{it} FD}s and $\mathcal{R}$(\SiII)}
\label{sec:ls-evolution}

\begin{figure}   
   \centering
   \includegraphics[width=8.4cm]{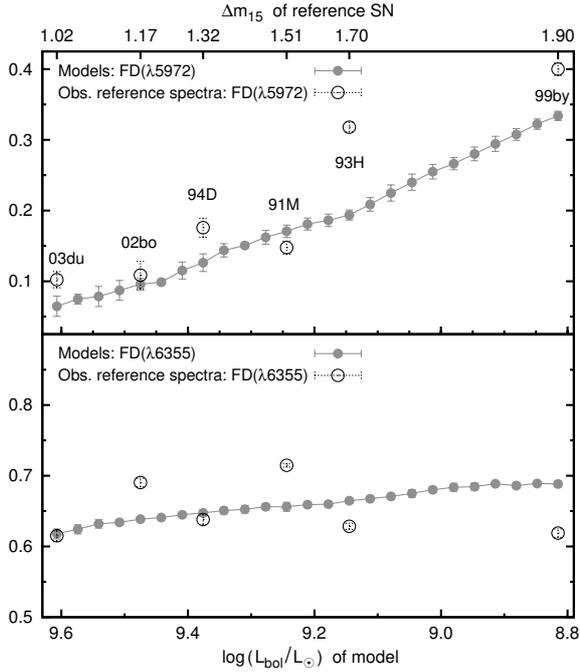}
   \caption{$\lambda 5972$ and $\lambda 6355$ {\usefont{OML}{cmm}{m}{it} FD} in models and observed reference spectra.} 
   \label{fig:rsi-contrib}
\end{figure}

\begin{figure}
   \centering
   \includegraphics[width=8.4cm]{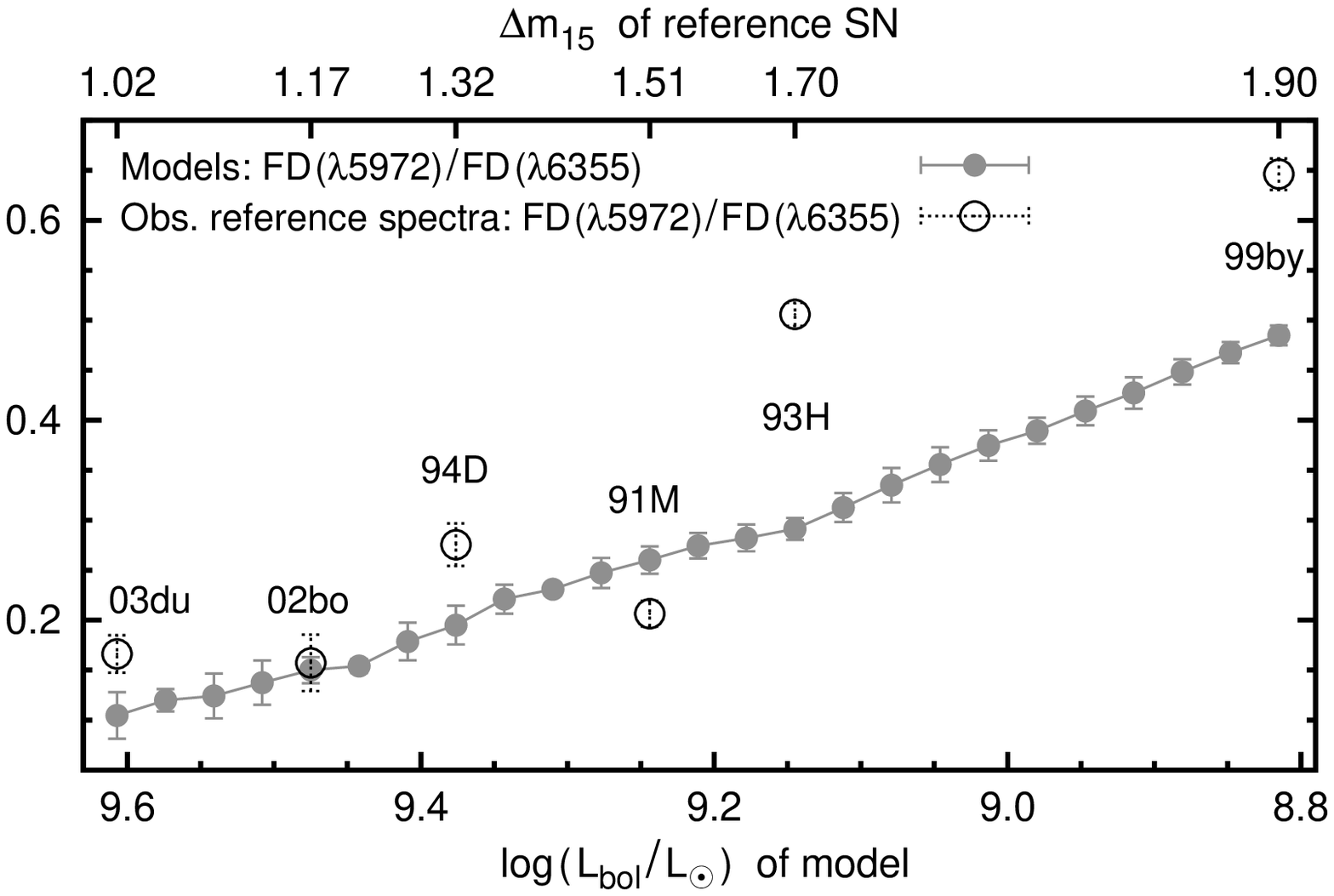}
   \caption{Ratio of \SiII\ $\lambda 5972$ to \SiII\ $\lambda 6355$ {\usefont{OML}{cmm}{m}{it} FD} [$\sim\!\!\mathcal{R}$(\SiII)].} 
   \label{fig:rsi}
\end{figure}

Fig. \ref{fig:rsi-contrib} shows our measurements of 
$\textrm{{\usefont{OML}{cmm}{m}{it} FD}}(\textrm{\SiII\ }\lambda5972)$ and
$\textrm{{\usefont{OML}{cmm}{m}{it} FD}}(\textrm{\SiII\ }\lambda6355)$.
While $\textrm{{\usefont{OML}{cmm}{m}{it} FD}}(\textrm{\SiII\ }\lambda5972)$ clearly
increases with decreasing luminosity, 
$\textrm{{\usefont{OML}{cmm}{m}{it} FD}}(\textrm{\SiII\ }\lambda6355)$ 
is approximately constant in all measurements. The resulting ratio
$\textrm{{\usefont{OML}{cmm}{m}{it} FD}}(\textrm{\SiII\ }
\lambda5972)/\textrm{{\usefont{OML}{cmm}{m}{it} FD}}(\textrm{\SiII\
}\lambda6355)\!\sim\!\mathcal{R}(\textrm{\SiII})$ is shown in Fig. \ref{fig:rsi}.
The value increases towards lower luminosities. 

Observed spectra are less well behaved than the synthetic ones, and show some
constant deviation at the dim end. This could be improved with more
sophisticated models using abundance stratification \citep[e.g.][]{ste05},
possibly including some contribution of \NaI\ D to the $\lambda5972$ feature in
dim SNe [See e.g. \citet{fil92}. \NaI\ would make too broad a contribution to
the $\lambda5972$ feature in our homogeneous models.]. Yet, the qualitative
evolution of the two \SiII\ features, as well as their different behaviour, is
correctly reproduced.

\section{Analysis}
\label{sec:data-abs}

Here we analyse the behaviour of $\mathcal{R}$(\SiII), which is rooted in the
evolution of \SiII\ $\lambda5972$ and \SiII\ $\lambda6355$. We show that the latter
line is easily saturated because it is very strong. \SiII\ $\lambda5972$, in contrast, 
is stronger in dimmer objects owing to a larger fraction of \SiII.

Furthermore, we investigate if \TiII\ lines have an influence on $\mathcal{R}$(\SiII), 
as proposed by \citet{gar04}.

\subsection{Quantifying the conditions for absorption in our models}
\label{sec:absfrac}

Absorption strengths are determined by intrinsic line strengths, abundances,
density, ionisation and excitation structure. For absorptions which are (mostly)
caused by a single ion, line formation can be conveniently analysed. Before
performing the analysis, we introduce our method and quantities important in
this context.

\subsubsection*{Region of the envelope relevant for absorption at $\lambda_{F\:\!\!D}$}

The reference wavelength of a {\usefont{OML}{cmm}{m}{it} FD},
$\lambda_{F\:\!\!D}$, is blueshifted with respect to the rest frame wavelength
$\lambda_\textrm{rest}$, because absorption occurs in material moving towards
the observer.  Therefore, as seen from the observer's point of view, the
processes relevant for a particular {\usefont{OML}{cmm}{m}{it} FD} take place
within a thin layer where the atoms have a line-of-sight (ray-parallel) velocity
$v_{||}=(1-\frac{\lambda_{F\:\!\!D}}{\lambda_\textrm{rest}})\cdot c$. In the
narrow-line limit (Sobolev approximation), with
$\boldsymbol{v}\propto\boldsymbol{r}$, this zone is an infinitely thin
``iso-$v_{||}$ plane'' perpendicular to the ray. 

The code calculates physical quantities describing absorption in each radial
shell. Thus, a weighted average over several shells is required to obtain values
corresponding to a measured {\usefont{OML}{cmm}{m}{it} FD} (and the respective
iso-$v_{||}$ plane). This is denoted by angle brackets ``$<...>$'' below. The
weight of each shell is determined by the number of photons impinging on the
absorption surfaces of the multiplet within the shell.

\subsubsection*{Sobolev optical depth and absorption fraction}

Both of the features in $\mathcal{R}$(\SiII) are doublets
($\lambda\lambda5958,5979$ and $\lambda\lambda6347,6371$, respectively). The
total fraction of photons absorbed (``absorption fraction'' $\mathfrak{A}$),
which is expected to correspond to {\usefont{OML}{cmm}{m}{it} FD} values, can
then be calculated as:

\begin{equation}
\mathfrak{A}:=\langle 1-\prod_l e^{-\tau_{l,k}}\rangle,
\end{equation}

where $\tau_{\textrm{l,k}}$ is the Sobolev optical depth for line $l$ in shell $k$,
defined as

\begin{equation}
\label{eq:taulk}
\tau_{l,k} = t\cdot \frac{h c}{4\pi} n_{i}\big|_k B_{ij} \left(1-\frac{g_i n_j}{g_j n_i}\Big|_k\right).
\end{equation}

Here, $n_i$ and $n_j$ are the number density of ions in the lower and upper level of
the line, $i$ and $j$, respectively, and $t$ is the time from explosion. All
other symbols have their usual meaning.

\subsubsection*{The influence of ionisation and occupation numbers}

Besides the intrinsic line strength $B_{ij}$, we identify two major influences on
$\tau_{l,k}$: First, the number density of \SiII,  $N_{\textrm{\SiII}}$, which depends
on the ionisation balance. Second, the level occupation numbers, which determine how
many of the ions are available for the transition. The influence of occupation
numbers can be encapsulated into an ``excitation factor'' (symbols as above):

\begin{equation}
\mathcal{E}_{l} := \langle\frac{n_{i}}{N_{\textrm{\SiII}}}\Big|_k
\left(1-\frac{g_i n_j}{g_j n_i}\Big|_k\right)\rangle.
\end{equation}

This contains the fraction of ions in the line's lower level and the correction
for stimulated emission, which usually varies relatively weakly.

\subsection{\SiII\ $\lambda 5972$ - the trend behind the $\mathcal{R}$(\SiII)-$L$ correlation}
\label{sec:si5972-abs}

\begin{figure}   
   \centering
   \includegraphics[width=8.1cm]{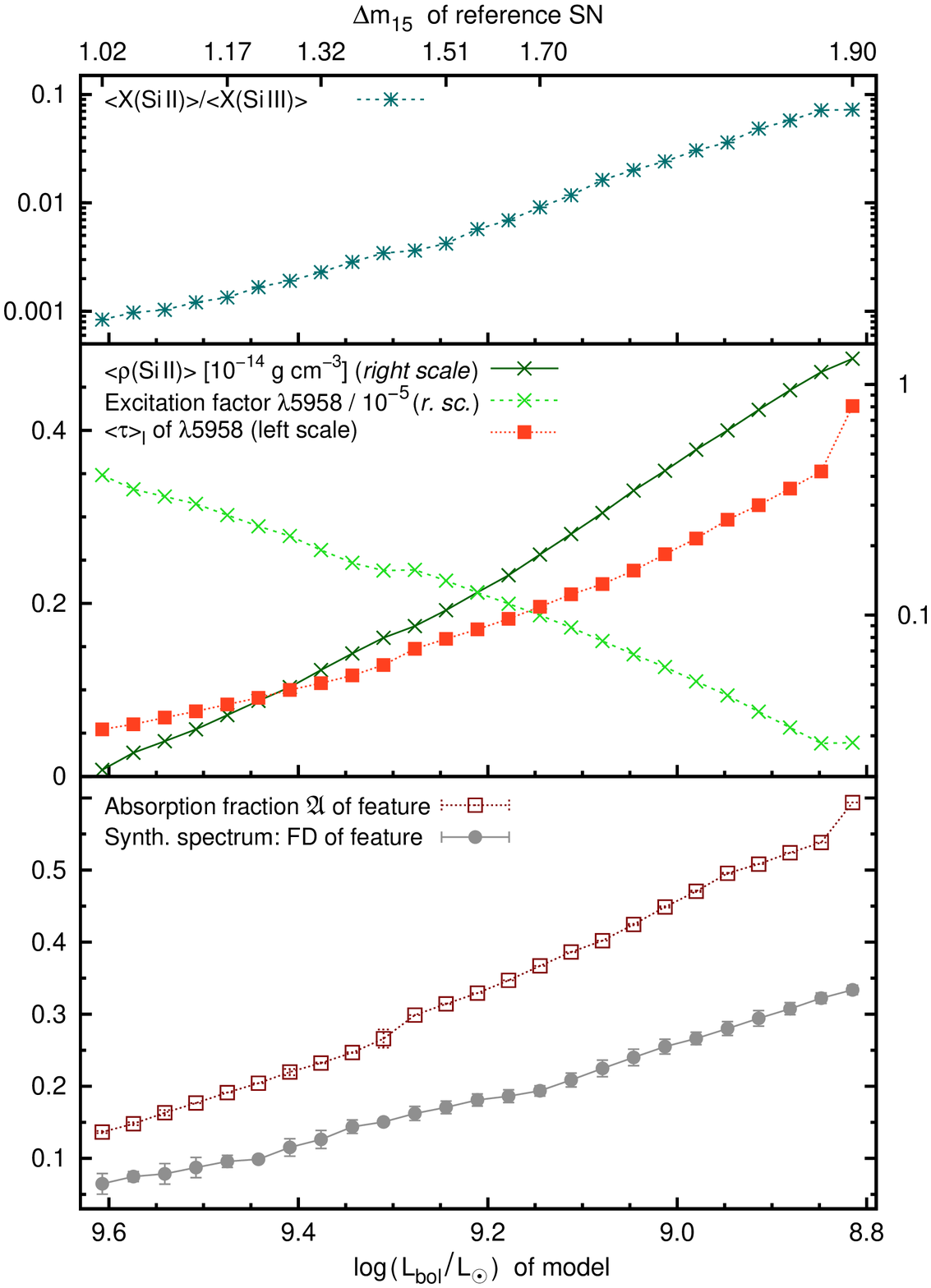}
   \caption{Absorption conditions for \SiII\ $\lambda 5972$.\protect\newline\textit{Upper panel:} Ratio of \SiII$\:\!$/$\:\!$\SiIII\ ionisation fractions averaged over the shells.\protect\newline\textit{Middle panel:} $\langle\rho($\SiII$)\rangle$, excitation factor $\mathcal{E}_l$ and the resulting optical depth $\langle\tau\rangle_l$. $\langle\tau\rangle_l$ is roughly determined by $\langle\rho($\SiII$)\rangle$ and $\mathcal{E}_l$ $\left(\textrm{crudely } \langle\tau\rangle_l \propto \langle\rho(\textrm{\SiII})\rangle\cdot\mathcal{E}_l \right)$. $\langle\tau\rangle_l$ and $\mathcal{E}_l$ are plotted for one representative line ($\lambda 5958$); the data for the other doublet line are approximately proportional.\protect\newline\textit{Lower panel:} Absorption fraction $\mathfrak{A}$, and measured {\usefont{OML}{cmm}{m}{it} FD}. Measurements in this and all following plots have been performed several times. Fluctuations in {\usefont{OML}{cmm}{m}{it} FD} and also in $\mathfrak{A}$ (due to fluctuations in $\lambda_{F\:\!\!D}$) are represented by error bars (whenever large enough).}
   \label{fig:si5972-abs}
\end{figure}

The \SiII\ $\lambda 5972$ absorption is stronger in dimmer SNe. This behaviour,
which is the basis for the $\mathcal{R}$(\SiII) sequence, is due to the \SiII\
to \SiIII\ ratio (Fig. \ref{fig:si5972-abs}, upper panel), which increases by
a factor of $\sim\!\!100$ from luminous to dim SNe. Other ionisation stages are
negligible in the layers probed.

The middle panel shows in detail the influence of the increasing \SiII\ density 
(crosses / dark, solid line) and of the excitation conditions on the optical depth.
The fraction of ions in the lower level of the line 
($\sim$ excitation factor, crosses / dashed line) decreases by 
$\sim\!\!90$\% towards lower luminosities and temperatures.
Yet, the optical depth of \SiII\ $\lambda 5972$ increases (filled squares 
/ dotted line).

The absorption fraction $\mathfrak{A}$ (lower panel, open squares / dark, dotted
line) increases towards lower luminosities, following the optical depth, as $
\mathfrak{A}=\langle 1-\prod_l e^{-\tau_{l,k}}\rangle\approx\langle\sum_l\tau_{l,k}\rangle$ 
for small optical depths (first order). Measured line {\usefont{OML}{cmm}{m}{it}
FD}s (circles / grey line) reflect this trend.

The offset between the exact values of {\usefont{OML}{cmm}{m}{it} FD} and the
absorption fraction, which increases somewhat towards the dim end, is discussed
in Appendix \ref{app:data-em}. As it turns out, $\sim\!\!50\%$ of this
offset is due to overlap with the emission part of the P-Cygni feature. The strength
of this generally correlates with the absorption.

\subsection{\SiII\ $\lambda 6355$ - a line with constant strength}
\label{sec:si6355-abs}

\begin{figure}   
   \centering
   \includegraphics[width=8.1cm]{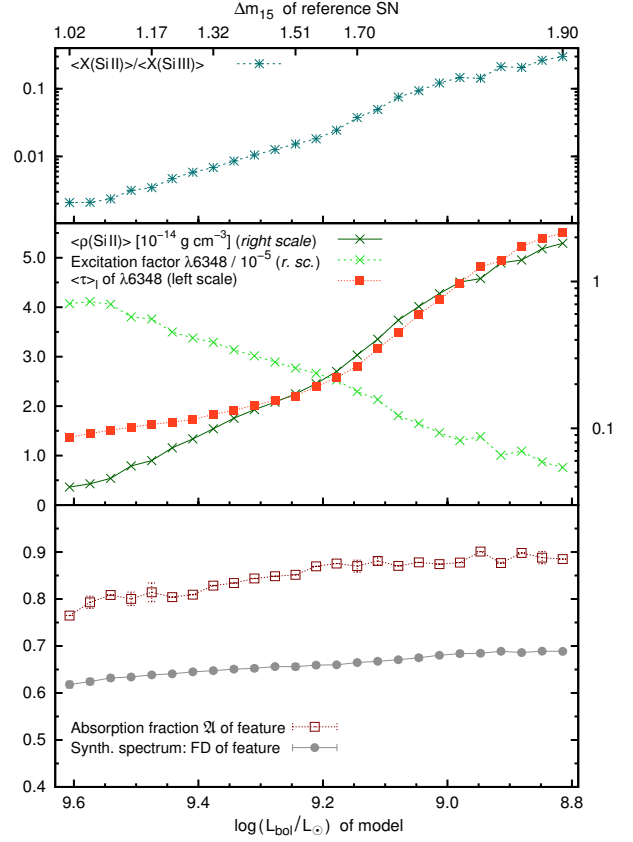}
   \caption{Absorption conditions for \SiII\ $\lambda 6355$; plot analogous to Fig. \ref{fig:si5972-abs}. The range of absorption fractions and {\usefont{OML}{cmm}{m}{it} FD}s is small because of saturation. Apart from intrinsic line strengths, the physics behind \SiII\ $\lambda 5972$ and \SiII\ $\lambda 6355$ is similar. Minor differences depend on different blueshifts of the features (see Appendix \ref{app:otherfeatures}, Fig. \ref{fig:fdvelocities}): \SiII\ $\lambda5972$ forms at $\gtrsim\!\!9000\ $\kms\ (luminous models) to $\sim\!\!8000\ $\kms\ (dim models); the stronger $\lambda 6355$ line always forms well above the photosphere at $\sim\!\!12500\ $\kms.}
   \label{fig:si6355-abs}
\end{figure}

\SiII\ $\lambda5972$ evolves with luminosity, while \SiII\ $\lambda6355$ does
not (Fig. \ref{fig:si6355-abs}, bottom panel). This is due to saturation:
because \SiII\ $\lambda6355$ has large optical depths (middle panel), the
absorption fraction (bottom panel) approaches unity, and it varies only weakly
($\mathfrak{A}=\langle 1-\prod_l e^{-\tau_{l,k}}\rangle\approx 1$ for large
$\tau$). Measured {\usefont{OML}{cmm}{m}{it} FD}s show a shallow trend, as
do absorption fractions. The {\usefont{OML}{cmm}{m}{it} FD} values indicate, however,
an additional amount of residual flux. Again, we show in Appendix \ref{app:data-em} that
50\% of this amount is due to the P-Cygni emission.

\subsection{Alternative explanations for $\textrm{{\usefont{OML}{cmm}{m}{it} FD}}(\textrm{\SiII\ }\lambda5972)$  and $\mathcal{R}$(\SiII)?}
\label{sec:additionalinfluences}

\begin{figure}
   \centering
   \includegraphics[angle=270,width=8.4cm]{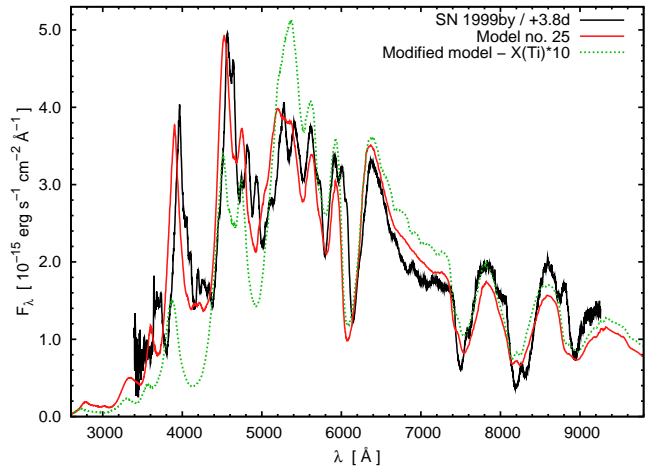}
   \caption{Model no. 25 and a modified version with 10-fold Ti abundance, where Ti lines contribute in the order of 10\% to the strength of the 5972\AA\ feature. A multitude of \TiII\ transitions is active, with strongest lines at $\lambda\lambda$5888.89, 5910.05, 5987.39, 5994.94, 6001.4, 6012.75. $\textrm{{\usefont{OML}{cmm}{m}{it} FD}}(\textrm{\SiII\ }\lambda5972)$ does not yet show an increase, also because \SiII\ $\lambda5972$ gets weaker due to the slightly modified temperature structure. The larger Ti abundance leads, however, to an unrealistic opacity increase below $\sim\!\!5200$\AA.}
   \label{fig:titest}
\end{figure}

\citet{gar04} suggested that the evolution of \SiII\ $\lambda 5972$ is due to a
contribution of \TiII\ lines in dim SNe. However, this possibility can be ruled
out reviewing the list of active lines. In order for \TiII\ lines to make a
significant contribution, the Ti abundances must increase by a factor of
$\gtrsim\!\!10$ in our SN~1999by model. This however leads to exceedingly large
strengths for many other Ti lines (Fig. \ref{fig:titest}). We also do not find
any further species that have a significant influence on the evolution of \SiII\
$\lambda\lambda5972,6355$, with the possible exception of \NaI\ in dim SNe (see
Sec. \ref{sec:ls-evolution}).

\section{Discussion}

\subsection{Spectroscopic luminosity indicators: $\cal R$(\SiII) and \SiII\ $\lambda 5972$ line strengths}

Having clarified the behaviour of observed {\usefont{OML}{cmm}{m}{it} FD}
values, we now briefly discuss the use of line strengths and ratios as
luminosity indicators. We have shown that {\usefont{OML}{cmm}{m}{it} FD}(\SiII\
$\lambda 5972$) correlates with luminosity and drives $\cal R$(\SiII). Thus,
{\usefont{OML}{cmm}{m}{it} FD}(\SiII\ $\lambda 5972$) itself (or similarly the
equivalent width {\usefont{OML}{cmm}{m}{it} EW}, as measured in
\citetalias{hac06}) should be the most direct and meaningful luminosity
indicator.

There are, however, uncertainties even in 
{\usefont{OML}{cmm}{m}{it} FD}(\SiII\ $\lambda 5972$)
as a spectroscopic luminosity indicator. First, the trend in the
line is due to the combination of trends that are nonlinear in temperature,
luminosity and also in \Dm\ in the relevant ranges (see Sec.
\ref{sec:data-abs}). In order to obtain luminosities from spectra, one has to
choose an appropriate functional form (e.g. linear or second order) to fit the
relation between feature strength and luminosity. Inaccuracies introduced by the
use of such a functional form can easily be overlooked or misinterpreted as
scatter. Second, intrinsic variations in the SNe and their spectra
limit the precision of spectroscopic luminosity indicators. Differences in
the abundance distributions among equally luminous SNe [see e.g.
\citet{tan08}, \citet{maz06}] are a possible reason for this (see also below,
Sec. \ref{sec:siphysics}).

One way to reduce the error would be to use a number of spectroscopic luminosity
indicators. Even quantities which are theoretically more elusive than 
{\usefont{OML}{cmm}{m}{it} FD} may be useful, because different indicators
respond differently to spectral peculiarities. Besides ratios like
$\mathcal{R}$(\SiII), {\usefont{OML}{cmm}{m}{it} EW} is also worth considering,
because it accounts better for absorption in the line wings. SNe for which
spectroscopic luminosity determination is unreliable could be identified if
different indicators gave inconsistent luminosities.

The accuracy of spectroscopic luminosity determination could be improved if
additional luminosity indicators were found that rely on species other than
\SiII. A number of such quantities has already been defined (\citealt{nug95},
\citealt{bon06}, \citetalias{hac06}, \citealt{fol08}). However, 
the validity of these potential luminosity indicators still 
needs to be critically assessed in terms of the
physical differences among SNe with different luminosities \citep{maz07}.

\subsection{{\usefont{OML}{cmm}{m}{it} FD}(\SiII\ $\lambda 5972$) and SN physics}
\label{sec:siphysics}

The \SiII\ $\lambda 5972$ optical depths, absorption fractions
and {\usefont{OML}{cmm}{m}{it} FD}s are quite tightly correlated. 
The line thus reflects the physics in the Si-dominated 
layers of most SNe Ia much better than \SiII\
$\lambda 6355$. It would be interesting to compare the
\SiII\ $\lambda 5972$ profiles in $B$ max. spectra of several 
SNe with similar luminosity and explore the differences.
If the Si distribution peaks in the zone with the largest \SiII\ ionisation 
fraction, one may expect relatively large {\usefont{OML}{cmm}{m}{it} FD}s. 
On the other hand, a smeared-out Si distribution, or one with several 
peaks might cause an unusually shallow feature, 
as observed in SN~2006X [\citealt{wan08}, \citealt{eli08} (in
prep.)].

Going one step further, it would be interesting to infer the distribution of Si
using stratified abundance models \citep{ste05,maz08}. This would be
especially valuable in the context of SN Ia theory (see \citealt{maz07}) and
explosion physics.

\subsection{Other features evolving with luminosity}

Apart from \SiII\ $\lambda 5972$ and $\lambda 6355$, other features in
photospheric spectra of SNe Ia seem to evolve with luminosity. The \SiII\
$\lambda 4130$ line behaves similar to \SiII\ $\lambda5972$ \citep{bro08}, and the
\SII\ trough is strongest for SNe with intermediate luminosity
\citepalias{hac06}. In Appendix \ref{app:otherfeatures}, these two features are
investigated.

\section{Conclusions}

Based on a sequence of 25 synthetic spectra constructed to fit SNe Ia with
different luminosities, we have been able to explain the
$\mathcal{R}($\SiII$)-L$ relation.

The increasing strength of \SiII\ $\lambda 5972$ is responsible for the increase
of $\mathcal{R}($\SiII$)$ from luminous to dim objects. This is in turn due to
ionisation: in dimmer SNe, \SiII\ $\lambda 5972$ is stronger because the \SiII\
fraction increases more strongly than than the level occupation numbers decline. In
contrast, \SiII\ $\lambda 6355$ is saturated because of its large optical
depths. Thus, the line has a rather constant strength and only a minor influence
on $\mathcal{R}($\SiII$)$.

An investigation as performed here shows how observed line strengths can be
related to SN physics. \SiII\ $\lambda 5972$ probably provides the most reliable
information about the Si layer in optical spectra.

If luminosity is to be inferred from spectra, a simultaneous evaluation of
several line strengths or ratios is most promising. In order to find additional
reliable luminosity indicators, one can apply our methods to further lines
caused by single ions.

\section*{ACKNOWLEDGEMENTS}
This work was supported in part by the European Community's Human Potential Programme under contract HPRN-CT-2002-00303, `The Physics of Type Ia Supernovae'. M.T. is supported through a JSPS (Japan Society for the Promotion of Science) Research Fellowship for Young Scientists. 

We thank the referee for his constructive comments. Furthermore we are grateful to everyone who provided us with spectra, especially V. Stanishev and M.~M. Phillips. SH thanks S. Taubenberger, especially for help with observational data.

We have made use of the NASA/IPAC Extragalactic Database (NED, operated by the Jet Propulsion Laboratory, California Institute of Technology, under contract with the National Aeronautics and Space Administration NASA). Furthermore, we used the \textsc{IRAF} (Image Reduction and Analysis Facility) software, distributed by the National Optical Astronomy Observatory (operated by AURA, Inc., under contract with the National Science Foundation), see http://iraf.noao.edu . \textsc{SYNPHOT} routines/data from \textsc{STSDAS} and \textsc{TABLES} v3.6 (products of the Space Telescope Science Institute, operated by AURA for NASA) have been used for synthetic photometry in \textsc{IRAF}.

\appendix

\section{The offset between fractional depths and absorption fractions}
\label{app:data-em}

While the general trends in \SiII\ $\lambda5972$ and $\lambda6355$, and thus in
$\mathcal{R}$(\SiII), can be explained by absorption effects, the mismatch
between $\mathfrak{A}$ and $F\!\!\:D$ deserves closer investigation. Here, we
analyse this offset.

\subsection{Influences on the fractional depth besides absorption}

\begin{figure}   
   \centering
   \includegraphics[width=8.3cm]{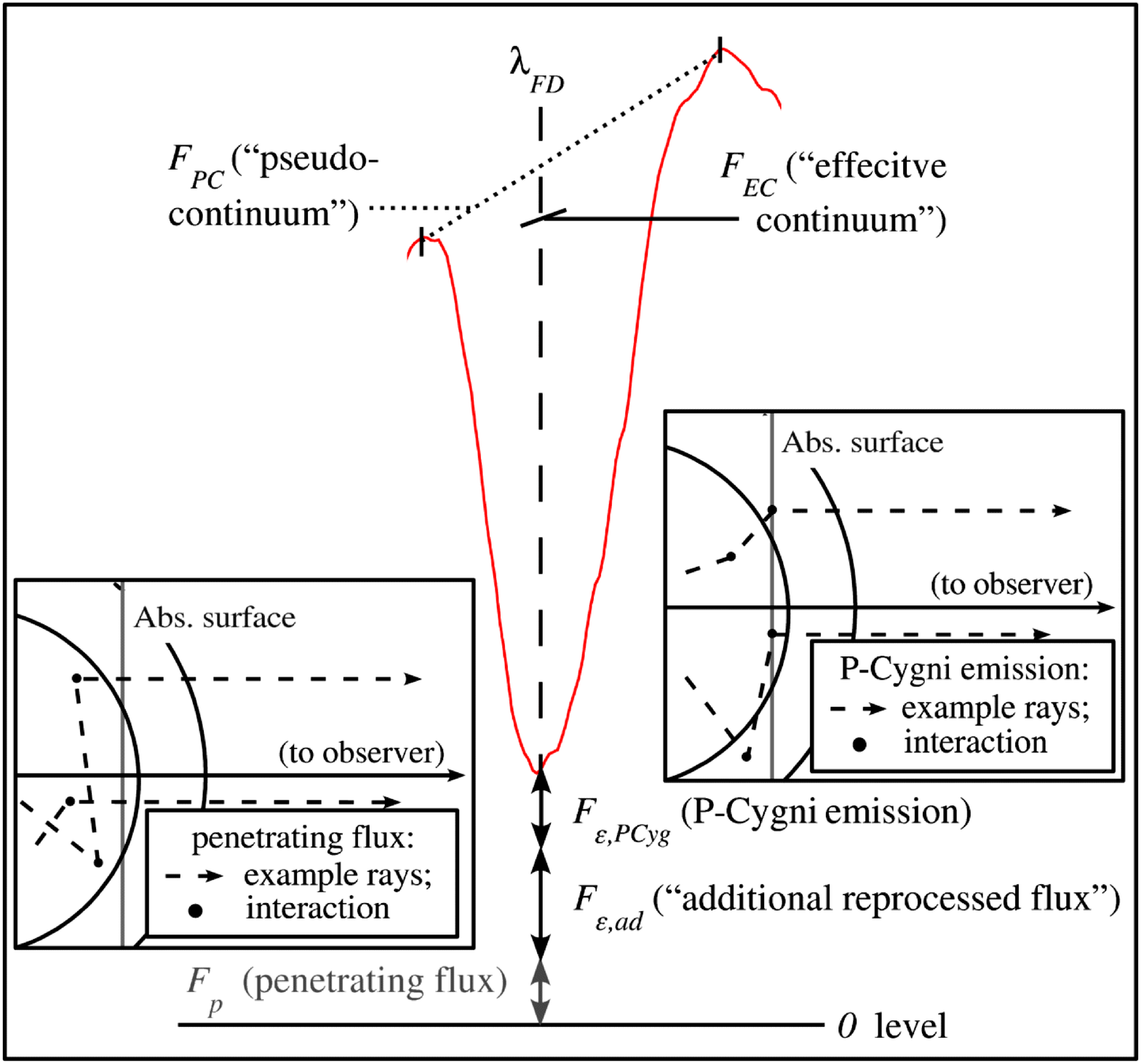}
   \caption{Contributions to the residual flux and continuum definitions. 
   Insets: Definition of $F_p$ (packets penetrating the absorption surface 
   without interaction) and of $F_{\epsilon,\textit{PCyg}}$ (packets having 
   interacted with the multiplet and reached the observer afterwards). 
   We term the remaining part of the emitted flux 
   $F_\epsilon = F_{\epsilon,\textit{PCyg}} + F_{\epsilon,\textit{ad}}$ 
   ``additional reprocessed flux'' ($F_{\epsilon,\textit{ad}}$).}
   \label{fig:line}
\end{figure}

There are two potential reasons for a difference between absorption fraction and
{\usefont{OML}{cmm}{m}{it} FD}: First, photons may be shifted into the observed
wavelength of a feature in re-emission processes, and enhance the residual
flux. Second, the estimated pseudo-continuum may deviate from the flux
behind the absorption surfaces, as seen from the observer. 

To trace these effects, we counted photon packets representing different
components of the residual flux in our simulations and recorded the
pseudo-continuum estimate for each {\usefont{OML}{cmm}{m}{it} FD} measurement.

\subsubsection*{Components of the residual flux}

The residual flux at a measurement wavelength $\lambda_{F\:\!\!D}$ can be
split into contributions of different origin (Fig. \ref{fig:line}).

One contribution (``penetrating flux'' $F_p$) directly depends on the 
absorption fraction $\mathfrak{A}$. It consists of photons crossing the 
multiplet's absorption surfaces without 
being absorbed, and is approximately given by
\begin{equation}
F_p \!\approx\! (1- \mathfrak{A}) F_{EC}
\end{equation}
where $F_{EC}$ (``effective continuum'') is the flux impinging on the 
absorption surfaces of the multiplet, multiplied by a correction factor. This factor,
usually a bit smaller than unity, accounts for the fact that photons
passing the absorption surface do not have a 100\% chance of escaping
the envelope.

The rest of the residual flux (``emission'' $F_{\epsilon}$) consists of photons
which have been shifted to an observer-frame wavelength $\sim\!\!\lambda_{F\:\!\!D}$
at the absorption surfaces (corresponding to $\lambda_{F\:\!\!D}$) or even
closer to the observer. In the following, photons that underwent absorption and
reemission at the absorption surfaces will be called ``P-Cygni emission''
($F_{\epsilon,\textit{PCyg}}$). The amount of P-Cygni emission
will generally correlate with the absorption strength. 
The flux of photons that were reprocessed into
an observer-frame wavelength $\sim\!\!\lambda_{F\:\!\!D}$ (for example in line
processes) at depths closer to the observer will be called ``additional
reprocessed flux'' ($F_{\epsilon,\textit{ad}}$). It is an upper limit estimate
for the flux that enters the feature without correlation to the absorption
conditions.

\subsubsection*{The relation between $\mathfrak{A}$ and {\usefont{OML}{cmm}{m}{it} FD}}

In general, the relation between absorption fraction, emission and measured 
residual flux $\frac{F_\textrm{res}}{F_{PC}}=1-FD$, is:
\begin{equation}
\label{eq:predictedres}
\frac{F_\textrm{res}}{F_{PC}} \equiv (1-FD) \!=\! \frac{F_{p}\! + \!F_{\epsilon}}{F_{PC}} \Bigg|_{\!\lambda_{\!F\:\!\!D}}\!\!\!\!\!\!\!\approx\! (1- \mathfrak{A}) \frac{F_{EC}}{F_{PC}} + \frac{F_{\epsilon}}{F_{PC}}.
\end{equation}

$\mathfrak{A}$ and {\usefont{OML}{cmm}{m}{it} FD} are closely related only if
$\frac{F_{EC}}{F_{PC}}$ and the additional reprocessed flux are roughly constant
with luminosity. As the data will show, this is the case for \SiII\ $\lambda 5972$
and $\lambda 6355$, but not always for the lines discussed in Appendix 
\ref{app:otherfeatures}.

\subsection{\SiII\ $\lambda 5972$}
\label{sec:si5972-em}

\begin{figure}
   \centering
   \includegraphics[angle=270,width=8.6cm]{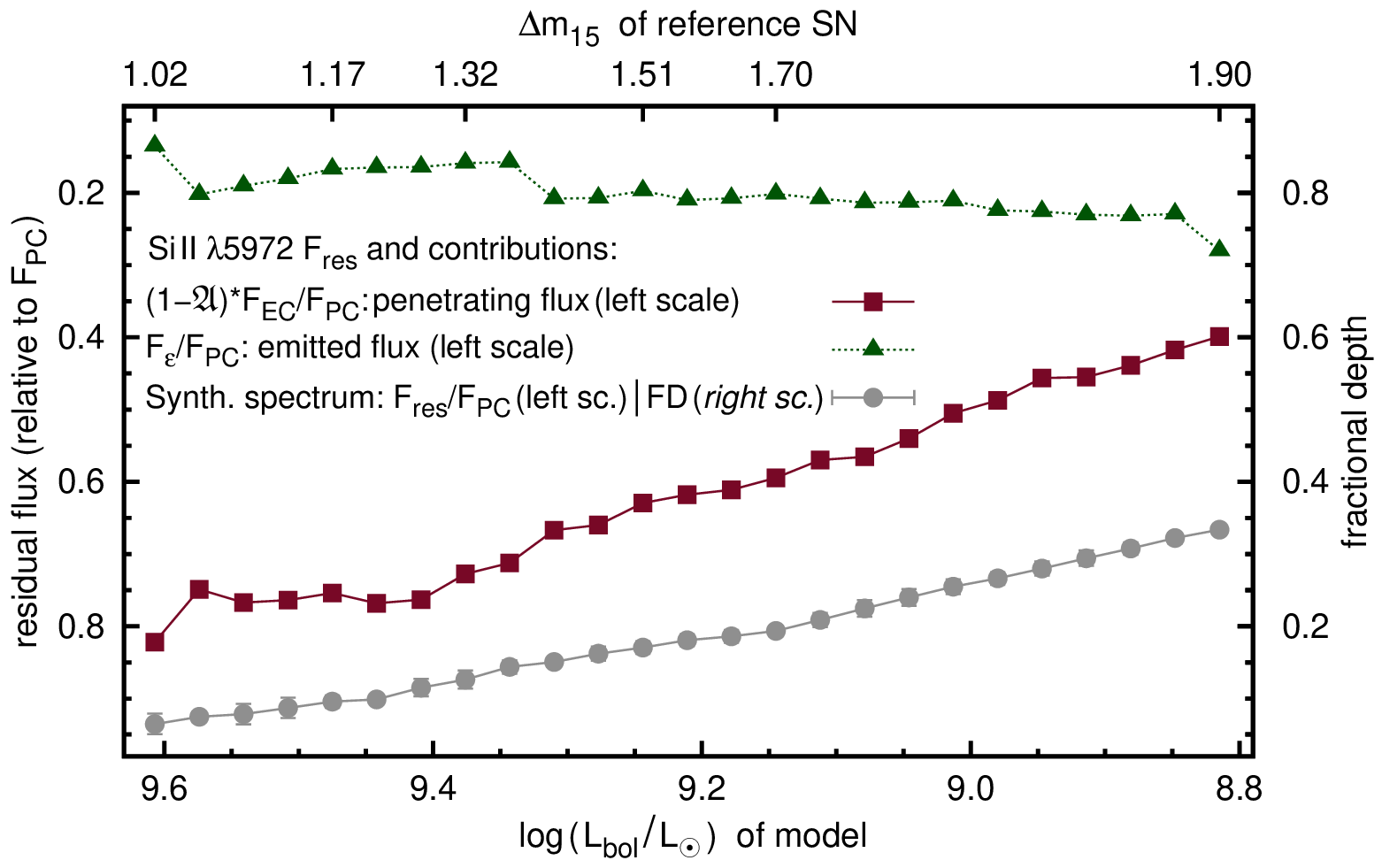}
   \caption{Contributions to the emission, and {\usefont{OML}{cmm}{m}{it} FD} for \SiII\ $\lambda 5972$. All fluxes are given relative to the pseudo-continuum. The penetrating flux  $(1- \mathfrak{A})\!\times\!\frac{F_{EC}}{F_{PC}}$ and the emitted flux $\frac{F_{\epsilon}}{F_{PC}}$ can be added up on the left scale to obtain a prediction for the residual flux (not explicitly shown). This coincides roughly with the measured residual flux (grey curve). The grey curve represents the fractional depth with units on the \textit{right hand side}.}
   \label{fig:si5972-em}
\end{figure}

For \SiII\ $\lambda 5972$, emission causes a moderate offset between penetrating
flux and measured residual flux $F_\textrm{res}=1-\textrm{\usefont{OML}{cmm}{m}{it} FD}$
(Fig. \ref{fig:si5972-em},
squares / dark line and circles / grey line, respectively). On average, about
half of the emission (triangles / dark, dotted line) is P-Cygni emission, which
increases with decreasing luminosity. The additional reprocessed flux, in turn,
is relatively constant (not explicitly shown), as is $\frac{F_{EC}}{F_{PC}}$. 
Thus, the measured {\usefont{OML}{cmm}{m}{it} FD}s correlate well with the absorption 
fractions.

Still, the offset between {\usefont{OML}{cmm}{m}{it} FD}s and absorption
fractions has a practical consequence: It makes the relative decrease in 
{\usefont{OML}{cmm}{m}{it} FD} from intermediate to very luminous SNe 
larger, increasing the influence of \SiII\ $\lambda 5972$ on $\mathcal{R}$(\SiII).

\subsection{\SiII\ $\lambda 6355$}
\label{sec:si6355-em}

\begin{figure}   
   \centering
   \includegraphics[angle=270,width=8.6cm]{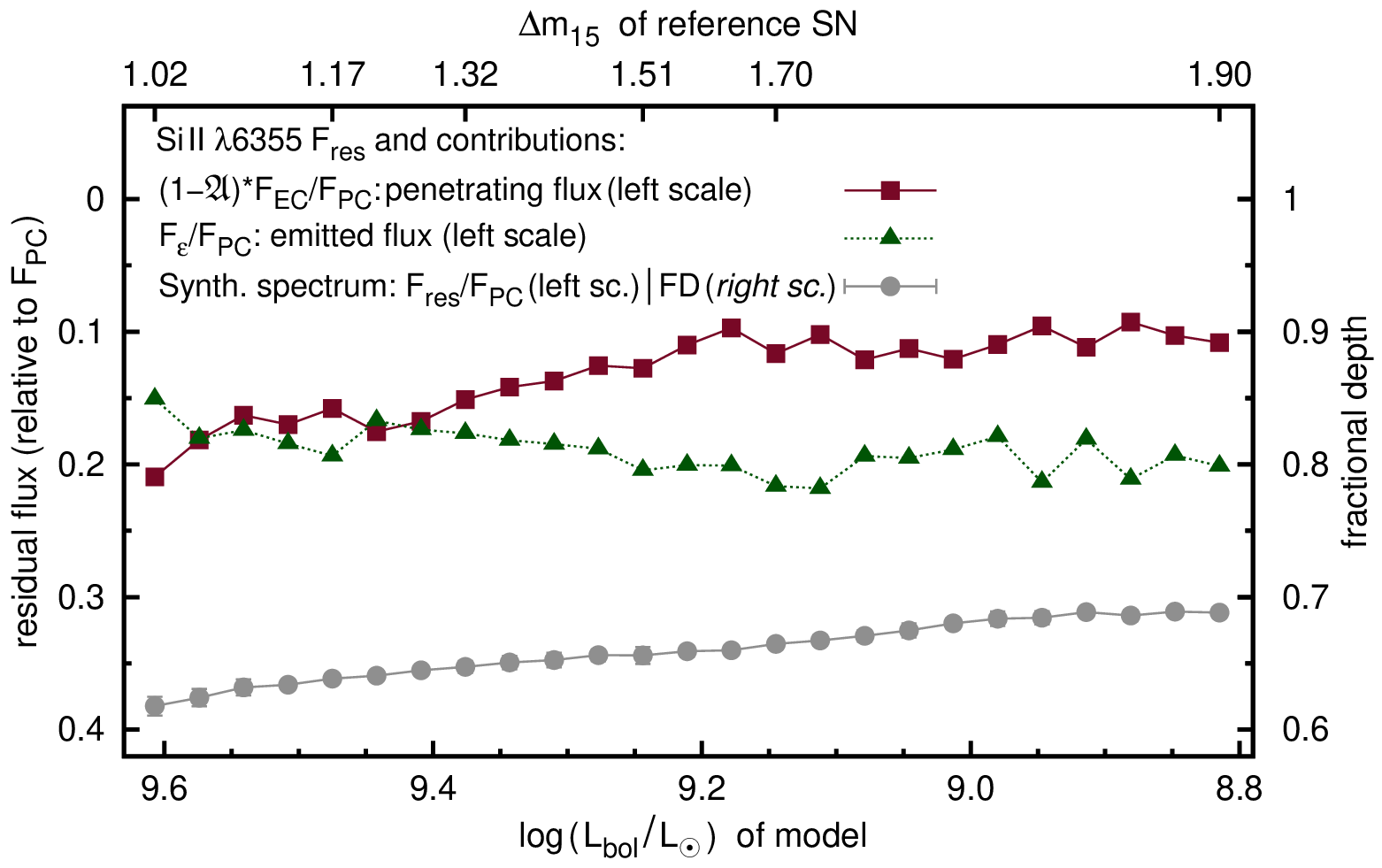}
   \caption{Emission contributions and {\usefont{OML}{cmm}{m}{it} FD} for \SiII\ $\lambda 6355$; plot analogous to Fig. \ref{fig:si5972-em}.}
   \label{fig:si6355-em}
\end{figure}

Intuitively, the residual flux in the saturated \SiII\ $\lambda 6355$ line
should be lower than observed. Again, this is in roughly equal parts due to 
P-Cygni emission and the additional reprocessed flux. Both these contributions
and $\frac{F_{EC}}{F_{PC}}$ are practically constant within the luminosity sequence. 

\subsection{Concluding remarks}

For the two lines determining $\mathcal{R}$(\SiII), $\frac{F_{EC}}{F_{PC}}$ is constantly
$\approx\! 1$ and the additional reprocessed flux is small ($\lesssim\!\!10\%$ of the residual flux).
Even under slightly different simulation conditions, we expect {\usefont{OML}{cmm}{m}{it} FD}s 
to correlate well with the absorption in their respective lines, because redwards of \SiII\ 
$\lambda 5972$, the number of lines which might cause deviations is relatively small.

\section{Investigation of further features (Si\textsc{ II} 4130\AA, S\textsc{ II} 5640\AA)}
\label{app:otherfeatures}

SN Ia spectra contain a wealth of features, of some of which show evolution
with luminosity (see e.g. \citealt{nug95} and \citetalias{hac06}). 
Calcium absorptions frequently seem to suffer from saturation and 
high-velocity contributions \citep{hat99,maz05,tan06}.
Interesting trends have been found for \SiII\ $\lambda 4130$ and the \SII\
``W-trough'': \SiII\ $\lambda 4130$ correlates with luminosity like \SiII\
$\lambda 5972$ \citep{bro08}, although the \SiII\ $\lambda 4130$-luminosity
correlation suffers from more scatter (\citealt{bro08}, Fig. 10). The \SII\
trough, in contrast, shows a parabolic trend, with largest strength at
intermediate luminosities \citepalias{hac06}. Both features are largely caused
by a single ion and can be investigated with the technique applied to
$\mathcal{R}$(\SiII).

\subsection{\SiII\ $\lambda 4130$: a supplementary luminosity indicator}

\begin{figure}   
   \centering
   \includegraphics[width=8.1cm]{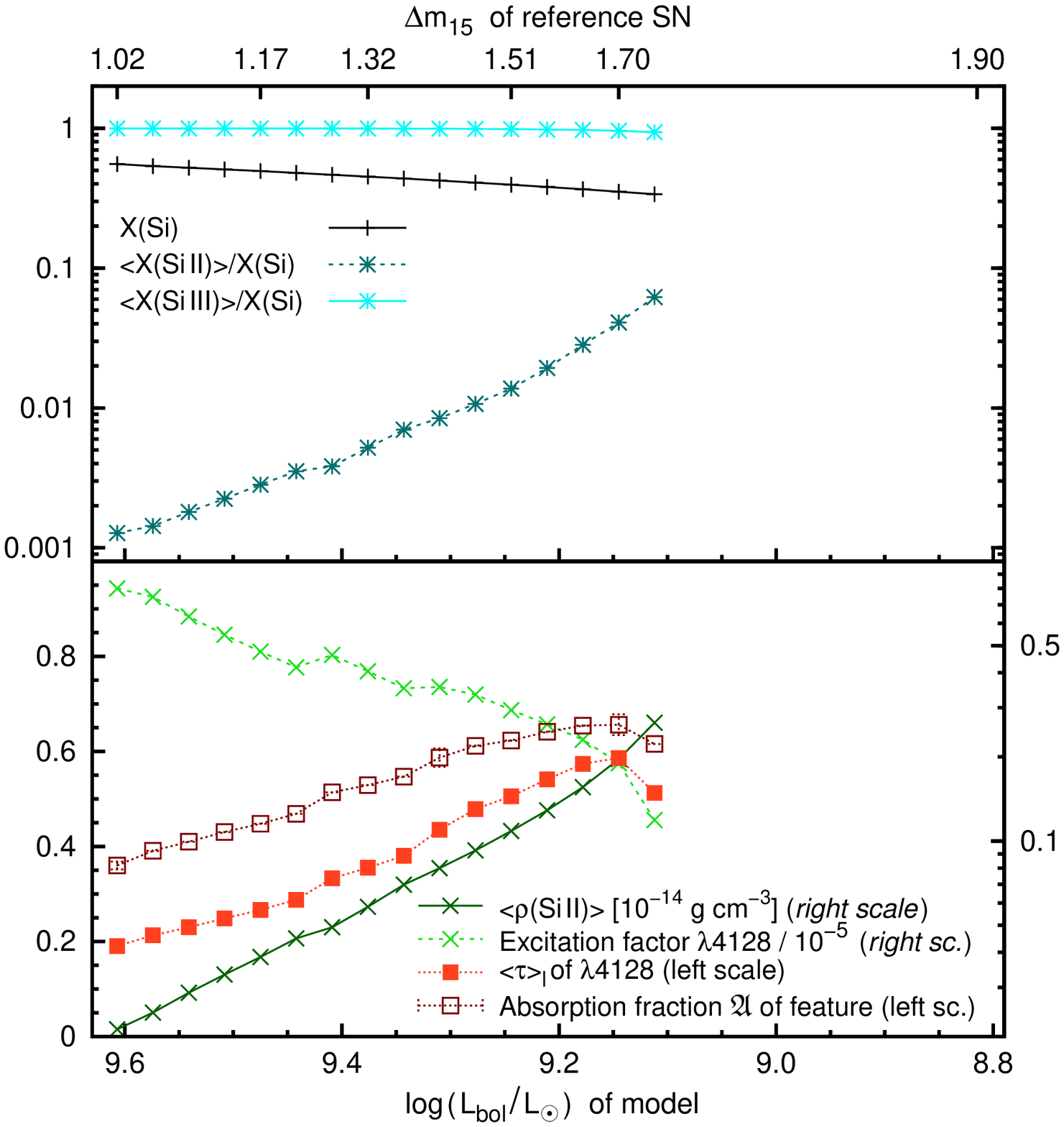}
   \caption{Absorption conditions for \SiII\ $\lambda 4130$ (for 
   $\log(L_\textrm{bol}/\Lsun)\gtrsim\!\!9.1$, see text).
   \protect\newline\textit{Upper panel:} Ionisation and abundance 
   trends: Si mass fraction, and \SiII\ and \SiIII\ ionisation fractions 
   (shell-averaged).\protect\newline\textit{Lower panel:} 
   $\langle\rho($\SiII$)\rangle$ and excitation factor $\mathcal{E}_l$; 
   $\langle\tau\rangle_l$ and absorption fraction $\mathfrak{A}$. 
   $\langle\tau\rangle_l$ and $\mathcal{E}_l$ are plotted for one representative line ($\lambda 4128$); the data for the other triplet lines are approximately proportional.}
   \label{fig:si4130-abs}
\end{figure}

\begin{figure}   
   \centering
   \includegraphics[angle=270,width=8.6cm]{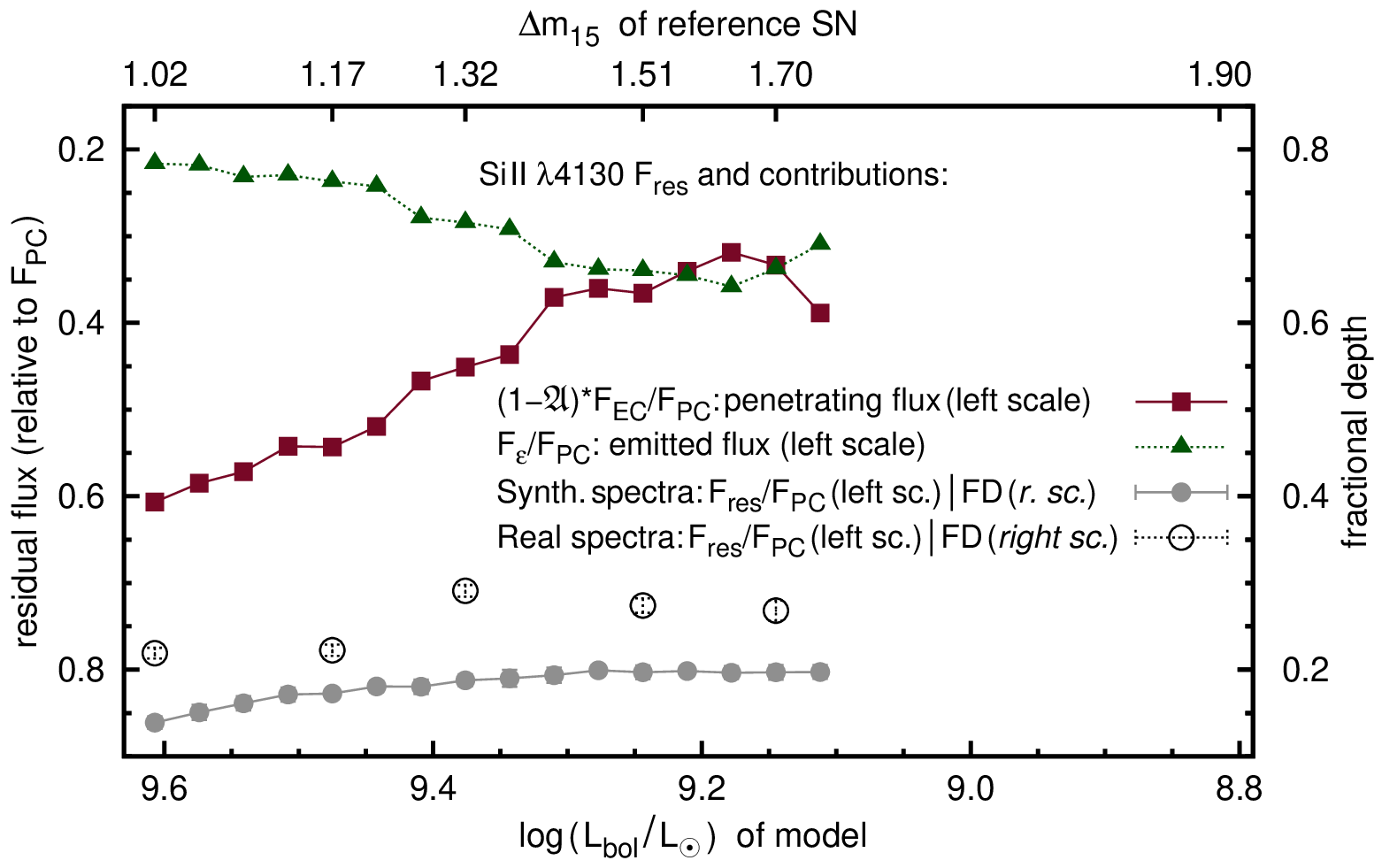}
   \caption{Contributions to the emission, and measured residual flux (grey curve) for \SiII\ $\lambda 4130$. 
    The grey curve represents the fractional depth with units on the \textit{right hand side}.
    Measurements of the residual flux in the observed reference spectra are shown for comparison.}
   \label{fig:si4130-em}
\end{figure}

The data for this feature are summarised in Figs. \ref{fig:si4130-abs} and
\ref{fig:si4130-em}. The line can be measured only in spectra with
$\log(L/\Lsun)\!\!\gtrsim\!\!9.1$. At lower luminosities, the \TiII\ lines
at $\sim$4000-4300\AA\ typical of dim SNe affect the feature.

Among normal SNe, the absorption fraction shows a trend similar to \SiII\ $\lambda 5972$
(Fig. \ref{fig:si4130-abs}, lower panel). Measured {\usefont{OML}{cmm}{m}{it}
FD} values (Fig. \ref{fig:si4130-em}, circles) correlate with the absorption
fractions, but increase only slowly towards lower luminosities. This is mainly due to 
an increase in the additional reprocessed flux (not explicitly shown), which leads to an increase
in the emission (triangles / dark, dotted line). Contamination effects 
(e.g. by \SII\ $\lambda 4162.7$, \NiII\ $\lambda 4067.0$ and \CoII\ $\lambda 4160.7$) are
a possible reason for this.

As noted by \citet{bro08}, the \SiII\ $\lambda 4130$ strength could be used as a
luminosity indicator for intermediate to bright objects. However, 
{\usefont{OML}{cmm}{m}{it} FD} does not follow the absorption fraction very well
for the feature. Thus, the relation between luminosity and \SiII\ $\lambda 4130$
line strength shows large scatter [\citet{bro08}, Fig. 10], and we regard the
feature as a supplementary luminosity indicator at most.

\subsection{\SII\ $\lambda 5640$: a mix of temperature and abundance trends}

\begin{figure}   
   \centering
   \includegraphics[width=8.1cm]{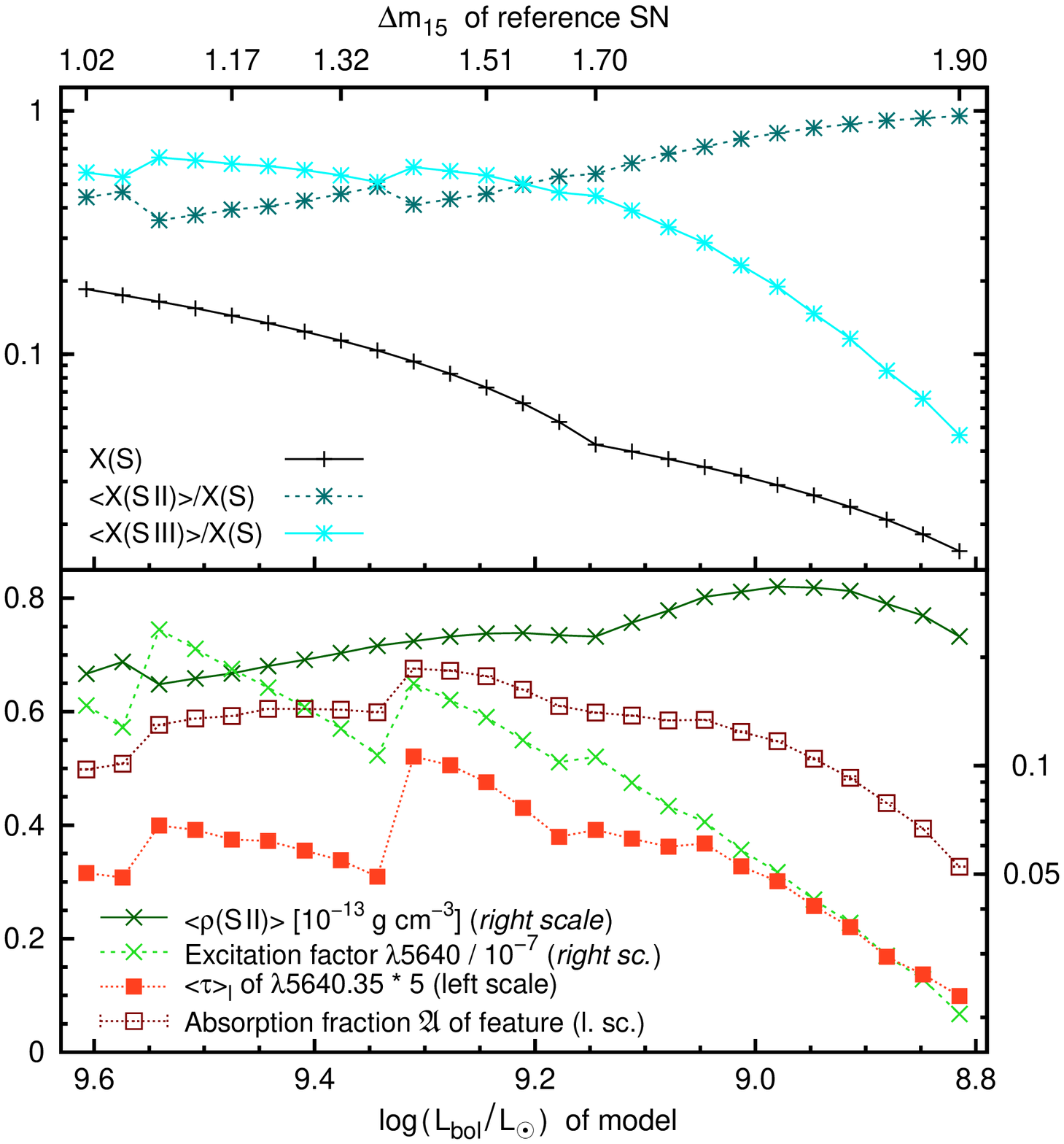}
   \caption{Absorption conditions for \SII\ $\lambda 5640$ (red part of W-trough). Abrupt variations in the values are due to jumps in the measurement positions (see Fig. \ref{fig:fdvelocities}). \protect\newline\textit{Upper panel:} S mass fraction of our models, and \SII\ and \SIII\ ionisation fractions (shell-averaged). \protect\newline\textit{Lower panel:} $\langle\rho($\SII$)\rangle$ and excitation factor $\mathcal{E}_l$; $\langle\tau\rangle_l$ and absorption fraction $\mathfrak{A}$. $\langle\tau\rangle_l$ and $\mathcal{E}_l$ are plotted for one example line (lower level 13.7eV, non-metastable). The lines of the \SII\ $\lambda 5640$ multiplet originate from different levels at $\sim$13.7...14eV, including a metastable one. Trends in $\langle\tau\rangle_l$ values for other lines will thus deviate somewhat, causing the absorption fraction to deviate from the individual optical depth trends.}
   \label{fig:s5640-abs}
\end{figure}

\begin{figure}   
   \centering
   \includegraphics[angle=270,width=8.6cm]{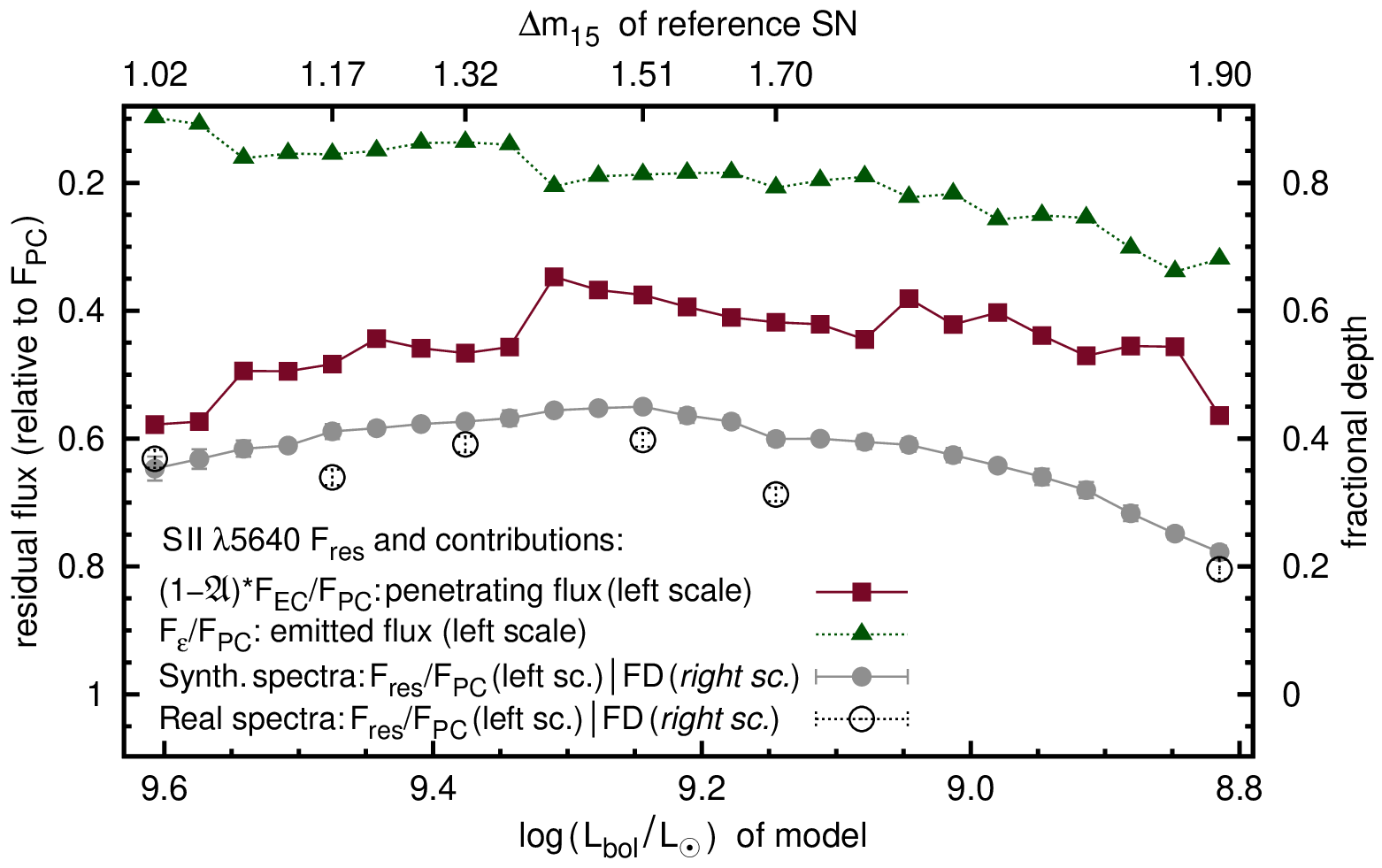}
   \caption{Emission contributions and {\usefont{OML}{cmm}{m}{it} FD} for \SII\ $\lambda 5640$. Plot analogous to Fig. \ref{fig:si4130-em}.}
   \label{fig:s5640-em}
\end{figure}

\begin{figure}
   \centering
   \includegraphics[angle=270,width=8.4cm]{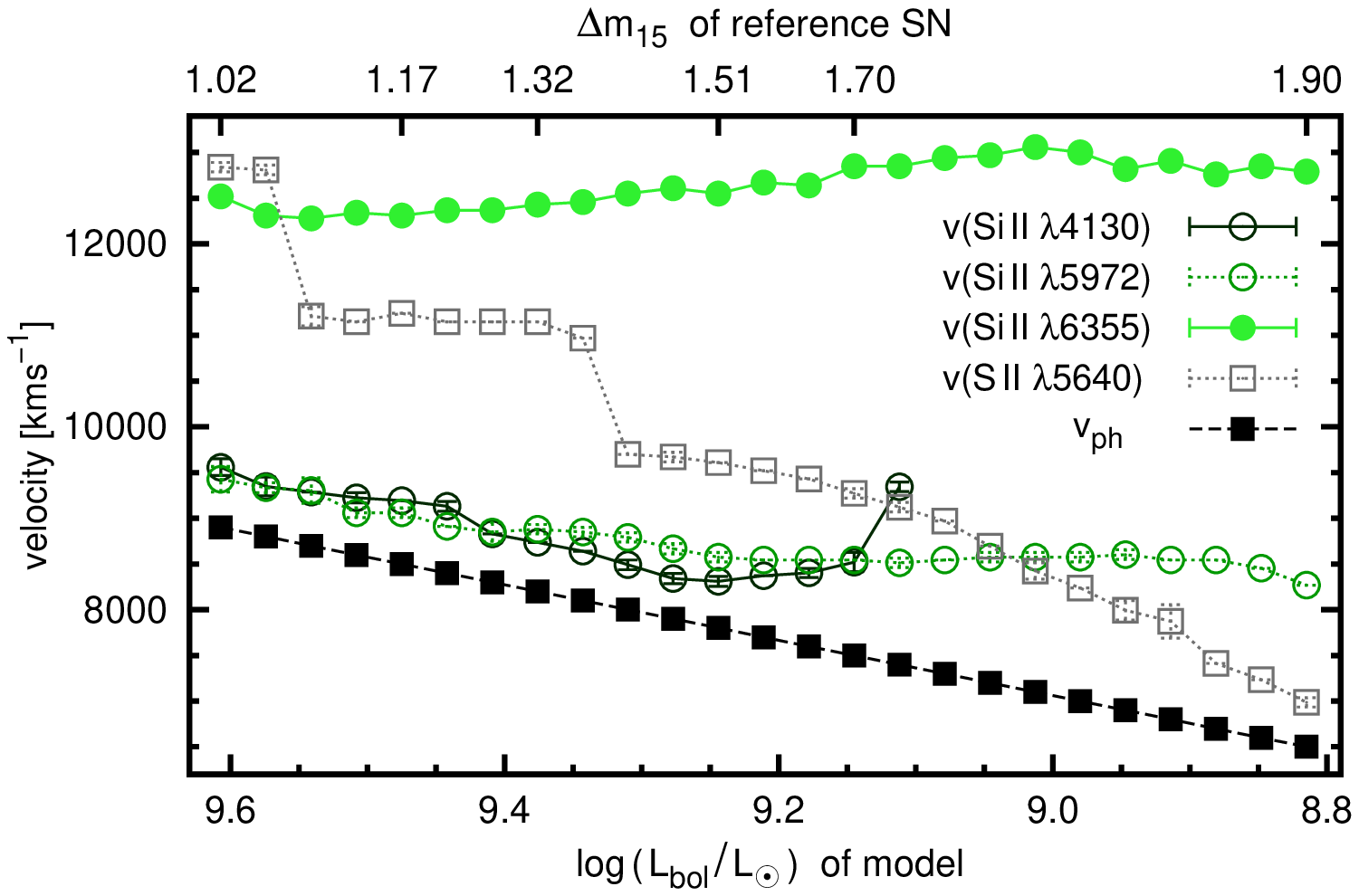}
   \caption{Line-of-sight velocities corresponding to $\lambda_{F\:\!\!D}$ values (in the synthetic spectra) for all features discussed in this paper. Weaker features tend to form near the photosphere at lower velocities, and trace the photospheric recession (filled squares / black, dashed line).}
   \label{fig:fdvelocities}
\end{figure}

The blue part of the \SII\ W-trough has different depths in the synthetic and
observed spectra (Fig. \ref{fig:spectra}, see also \citealt{kot05}). Thus, we
chose to investigate only the red component (lines/multiplets $\lambda\lambda
5606, 5640$). The pseudo-continuum was set as if the whole trough were
measured. 

The results are shown in Figs.~\ref{fig:s5640-abs} and \ref{fig:s5640-em}.
Sulphur has different level structures {and a lower ionisation threshold than
silicon, so that the \SII/\SIII\ ratio is always $>\!\!0.5$.}
Furthermore, because of the steeply declining line velocity (Fig.
\ref{fig:fdvelocities}), the temperatures in the layers probed do not vary as
much as for the Si lines within the sequence.

In the range from large to intermediate SN luminosity, a mild increase in the
\SII\ fraction (Fig. \ref{fig:s5640-abs}, upper panel, asterisks / dark, dashed
line) increases the optical depths somewhat. At low luminosities, the \SII\
fraction approaches 100\%, so that decreasing S abundances (Fig.
\ref{fig:s5640-abs}, upper panel, crosshairs / solid line) and level occupation
numbers lead to a decreasing absorption strength (Fig. \ref{fig:s5640-abs},
lower panel).

The resulting {\usefont{OML}{cmm}{m}{it} FD}s (Fig. \ref{fig:s5640-em}, circles)
in the synthetic spectra increase from large to intermediate luminosities and
then decline again, similar to the parabolic trend in \citetalias{hac06} for the
$EW$ of the W-trough. The decline in {\usefont{OML}{cmm}{m}{it} FD} from
intermediate- to low-luminosity SNe is amplified by an increase in the
additional reprocessed flux, which affects dim SNe more strongly (triangles / dark,
dotted line). 

\end{document}